\documentclass{article}
\usepackage{graphicx}
\usepackage{color}
\usepackage{epsfig}
\usepackage{amssymb}
\usepackage{amsmath}

\usepackage{algorithm}
\usepackage{algpseudocode}

\newtheorem{theorem}{\bf Theorem}
\newtheorem{result}{\bf Basic Result}

\newcommand{\proof}[1]{\textbf{\textit{Proof:}}  #1  \hfill\colorbox{black}{}}

\begin{document}


\title{MAINT: Localization of Mobile Sensors with Energy Control}

\author{B. Sau \\ \small
    Department of Mathematics\\ \small Jadavpur University, Kolkata, India\\
\footnotesize    buddhadeb.sau@gmail.com
    \and S. Mukhopadhyaya\\
   \small Birla Institute of Technology\\ \small Mesra, India\\
 \small srabanim@gmail.com
         \and
         K. Mukhopadhyaya \\
\small         Advanced Advanced Computing and Microelectronics Unit\\
\small          Indian Statistical Institute,       Kolkata, India\\
        \small krishnendu@isical.ac.in
}

\date{}
 \maketitle

\begin{abstract}
Localization is an important issue for Wireless Sensor Networks (WSN). A mobile sensor may change
its position rapidly and thus require localization calls frequently. A localization may require
network wide information and increase traffic over the network. It dissipates valuable energy for
message communication. Thus localization is very costly. The control of the number of localization
calls may save energy consumption, as it is rather expensive. To reduce the frequency of
localization calls for a mobile sensor, we propose a technique that involves \textit{Mobility Aware
Interpolation} (MAINT) for position estimation. It controls the number of localizations which gives
much better result than the existing localization control schemes using mobility aware
extrapolation. The proposed method involves very low arithmetic computation overheads. We find
analytical expressions for the expected error in position estimation. A parameter, the time
interval, has been introduced to externally control the energy dissipation. Simulation studies are
carried out to compare the performances of the proposed method with some existing localization
control schemes as well as the theoretical results. The simulation results shows that the expected
error at any point of time may be computed from this expression. 
We have seen that constant error
limit can  be maintained increasing the time period of localization proportional to rate of change
of direction  of its motion.  Increasing time period, the  energy may be saved with a stable error
limit.
\end{abstract}

\textit{keywords:}
Wireless sensor networks, localization, tracking mobile sensors, localization control, target 
tracking


\section{Introduction} \label{introduction}
A micro-sensor (or simply a sensor) is a small sized and low powered electronic device with limited
computational and communication capabilities. A WSN may contain some ten to millions of such
sensors. If the sensors are deployed randomly, or the sensors move about after deployment,
finding the locations of sensors (\textit{localization}) is an important issue in WSN. Localization
requires communication of several necessary information between sensors over the network and a lot
of computations. All these comes at the cost of high energy consumption. So far, research have
mainly been focused on finding efficient localization techniques in static sensor networks (where
the sensor nodes do not change their positions after deployment) \cite{BHE00,MKQP01,RSPS02}.

In a WSN, sensors may be deployed either by some design with predefined infrastructure or through
random manual placements of sensors. After being deployed, the sensors may remain static or move
with time. In both the cases, the positions of the sensors need to be determined. Bulusu et
al.~\cite{BHE00} proposed localization technique without using GPS. The techniques for finding
locations of sensors in static networks are costly. As sensors in mobile WSN change their positions
frequently, many localization calls are necessary to track a mobile sensor. A fast mobile sensor may
require frequent localizations, draining the valuable energy quickly. To reduce the number of
localization calls, positions of sensors in different time instant can be predicted or estimated
from the history of the path of the sensor \cite{BM02,HE04}.

Dynamic sensor networks have immense applications giving assistance to mobile soldiers in a battle
field, health monitoring, in wild-life tracking \cite{JOWMPR02}, etc. A moving sensor needs to  find
its position frequently.  Using GPS may not be appropriate due to its low accuracy, high energy
consumption, cost and size. An optimized localization technique of static sensor network is used to
find the current position of a mobile sensor.

Tilak et al \cite{TKAK05} proposed some techniques for tracking mobile sensors based on \textit{dead
reckoning} to control the number of costly localization operations. Among these techniques, the best
performance is achieved by \textit{MADRD}. It estimates the position of a sensor, in stead of
localizing the sensor every time it moves. Error in the estimated position grows with time. Every
time localization is called, the error in the estimated position is calculated. Depending on the
value of this error the time for the next localization is fixed. Fast mobile sensors trigger
localization with higher frequency for a given level of accuracy in position estimation. We
proposed a technique to estimate positions of mobile sensors with a control on localization calls
and with lower energy dissipation.

\textit{The main focus of this paper is as follows:}
In this paper, a method is proposed to estimate the positions of a mobile sensor, in stead of
localizing every time when its position is required. The proposed method estimates the position of a
sensor only when it is required by a base station. By this algorithm with a slight modification, a
mobile sensor may find its locations locally (i.e., distributively) rather than centrally in a base
station. The information of an inactive sensor is ceased to be communicated. Most calculations are
carried out at the base station to reduce arithmetic complexity of sensors. Localizations are called
with a time interval, $T$. In this paper, we consider that the sensors moves with the Random
Waypoint Mobility Model (RWP). We have seen that energy consumption may be regulate with the
parameter $T$. An analytical expression for expected error in position estimation are deduced. It
helps to fix the value of $T$ to regulate the energy dissipation controlling the the number of
localization calls with a knowledge of rate of changes in the direction of path of a sensor
depending on the applications. The proposed method gives higher accuracy in estimation for a
particular energy cost and vice versa. Both the analytical formula and simulation studies show that
our proposed algorithm incurs significantly lower error than that of MADRD even consuming equal
energy. Some part of this paper was published in a conference paper~\cite{SMM06}.

In the rest of the paper, Section~\ref{prob:stmt} describes the problem for tracking mobile sensor.
In Section~\ref{early:wrk} we discuss related works as well as our motivation to propose an
estimation method using interpolation. Section~\ref{protocol} describes the proposed algorithm for
tracking mobile sensors. Section~\ref{analysis} deals with the analysis of the algorithm and
different advantages. In Section~\ref{sim:res} simulation results are presented. Finally, we present
our conclusion in Section~\ref{conclude}.


\section{Problem Statement and Performance Measures} \label{prob:stmt} 

The position of a sensor is determined by a standard localization method. 
We assume that the location determined by this localization represents
the actual position of the sensor at that moment. The sensors are 
completely unaware of the mobility pattern. Therefore, the actual position  
of a sensor $S$ any time $t$ is unknown. The position may be estimated or 
found by localization call. The absolute error in location estimation may 
be calculated as:
   $$ error_{abs}= \sqrt{(x-\hat{x})^2+(y-\hat{y})^2}$$
where ~$(x,y)$  and $(\hat{x},\hat{y})$ denote the actual and
estimated positions at time $t$ respectively. Frequent calls for localization
consume enormous energy. To design an algorithm that optimizes both accuracy 
and energy dissipation simultaneously is very difficult. An efficient, 
robust and energy aware protocol is required to decide whether the 
location of the sensor would be estimated with a desired level of
accuracy or found by localization with an acceptable level of energy cost.

\section{Related Works and Motivation of this Work} \label{early:wrk} 

Researchers have mainly focused their attention to discovering efficient methods of localization
technique in static sensor networks \cite{PCB00,SHS01,TFBD01}. Thurn et al \cite{TFBD01} proposed
probabilistic techniques using Monte Carlo localization (MCL) to estimate the location of mobile
robots with probabilistic knowledge on movement over predefined map. They used a small number of
seed nodes (nodes with known position), as \textit{beacon}s. These nodes have some extra hardware.
Hu et al \cite{HE04} introduced the sequential MCL to exploit the mobility without extra hardware.
WSNs generally remain embedded in an unmapped terrain and sensors have no control on their mobility.
To reduce the number of localization calls was used~\cite{BM02} for saving energy. The positions of
a mobile sensor at different time instant are estimated from the history of the path of the sensor.

Tilak et. al~\cite{TKAK05} tried to reduce the frequency of localizations for finding the position
of mobile sensors. They proposed techniques: 1) SFR (\textit{Static Fixed Rate}), 2) DVM
(\textit{Dynamic Velocity Monotonic}) and 3) MADRD (\textit{Mobility Aware Dead Reckoning Driven}).
SFR periodically calls some classical localization operation. In this protocol, at the time of
reporting an event to the base station the sensor sends its position obtained in last localization.
Therefore, localization operations are called unnecessarily when a sensor is not moving. On the
other hand, reported location may suffer a large error from the actual position in moment of
reporting the event. DVM adaptively calls some localization with the mobility of the sensors. In
DVM, localizations are called with greater frequency when the sensor moves fast and lower frequency
when it moves slowly in a straight line. A sensor with high mobility drains the energy quickly and
dies soon. If a sensor suddenly moves with very high speed from rest, then error in reported
location becomes very high. The third method, MADRD, predicts locations of a sensor from its motion
between last two localizations using extrapolation. In MADRD, every time when localization is called
the actual position is reported. If the expected error (the distance between reported position and
the position according to prediction) is compared to a {\it threshold error}, $E_{thresh}$,
(implementation dependent). If the expected error exceeds $E_{thresh}$, the position predictor
becomes erroneous quickly. Localization calls should be triggered with higher frequency. Again a
sensor with high speed calls localizations frequently.

\begin{figure}[h!]
 \centering
 \includegraphics[width=.8\textwidth]{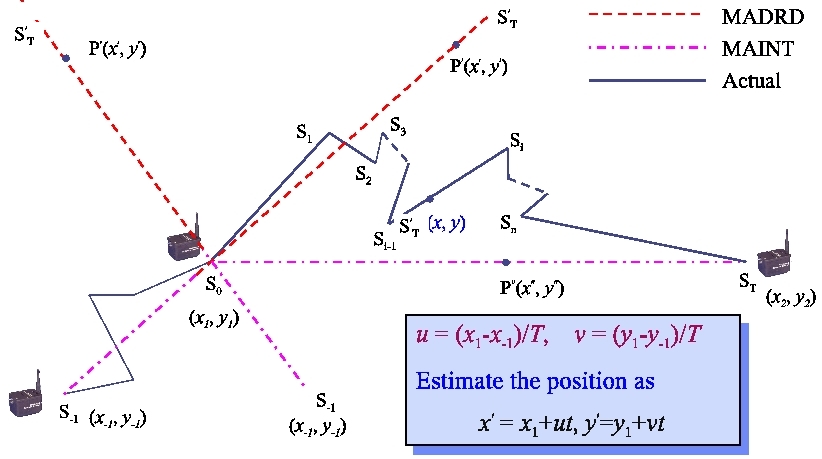}
 \caption{Figure showing the estimated path using interpolation and extrapolation.}
 \label{fig:motive}
\end{figure}

Our goal is to reduce the error consuming energy no more than that of MADRD. Figure~\ref{fig:motive}
shows that estimated path due to MADRD fluctuates depending on the actual motion in between last two
calls.
As opposed to MADRD, estimation using interpolation depends on localization calls enclosing the
point to be estimated rather than last calls. Our intuition is that estimation by interpolation will
be more guided by the actual motion than MADRD. If the sensors change the mobility pattern  (i.e.,
speed, direction of path etc.) very frequently MADRD incurs high error in the position estimation.
We propose the algorithm MAINT. We proved our intuition by deriving average error for MADRD as well
as MAINT. We support the analytical result by simulations. Both the analytical formula as well as
simulation studies show that our proposed algorithm incurs significantly lower error than that of
MADRD even consuming equal energy.


\section{Localization Protocol for Tracking Mobile Sensors} \label{protocol} 

In several applications, a mobile sensor may frequently change its position and direction of its
path of mobility with time. A simple strategy for finding its position is the use of standard
localization methods at any time. But if the position of the sensor is required frequently, this
method is very costly.
SFR calls a classical localization operation periodically with a fixed time interval. To respond a
query from the base station, a sensor sends its position obtained from the last localization. When a
sensor remains still or moves fast, in both cases, the reported position suffers a large error. In
DVM, localization is called adaptively with the mobility of the sensors. The time interval for the
next call for localization is calculated as the time required to traverse the \textit{threshold
distance} (a distance, traversed by the sensor, location estimation assumed to be error prone) with
the velocity of the sensor between last two points in the sequence of localization calls. In case of
high mobility, a sensor calls localization frequently. If a sensor suddenly moves with very high
speed from rest, error in the estimated location becomes very high. In MADRD, the velocity is
calculated from the information obtained from last two localized points. The predictor estimates the
position with this velocity and communicates to the query sender. At the localization point, the
localized position is reported to the query sender and the \textit{distance error} is calculated as
the distance between the predicted position and reported position.  If the error in position
estimation exceeds threshold error (application dependent), the predictor appears to be erroneous
and localization needs to be triggered more frequently. The calculation of error is necessary every
time a localization called. Also, a sensor with high speed calls localizations frequently. We have
proposed a method, MAINT, to estimate the current position with better trade off between the
energy consumption and accuracy. MAINT uses interpolation which gives better estimation in most
cases.

\subsection{Mobility Aware Interpolation (MAINT)}

In some applications, the base station may need the locations of individual sensors at different
times. The location may be required to be attached to the data gathered by the sensors in response
to a query. However, the data may not be required immediately. In such cases, the number of
localization calls may be reduced by delaying the response. We propose a localization control scheme
by estimating positions using interpolation. The sensor holds the queries requiring the the
location, into a list, {\tt queryPoints} and sends the event to the base station padding the time of
occurrence. At the following localization point, the sensor sends these two localized positions to
each of the query senders in the time interval between these two localization points which are
already in the list. The base station estimates the positions with more accuracy by interpolation
with this information. The time interval of localization calls is as simple as in SFR. It eliminates
all the arithmetic overheads as opposed to MADRD and the error prone nature in sudden change of
speeds. Unnecessary calls of localizations for slow sensors may be avoided. To reduce the energy
dissipation, the localization method may be called with higher time interval. The localization may
be called immediately after receiving the query for real time applications or some special purpose.
Each sensor runs a process described by Algorithm~\ref{algo:main}.
\begin{algorithm}[h!]
\caption{(MAINT: Proposed algorithm)}
\begin{algorithmic}[1]
\State Let $(x_1,y_1)$ denotes the last localization point occurred at time $t_1$.
\State Set $\mathtt{queryPoints} \gets \emptyset$.
\While{(a query received from a sensor $S$ at time $t > t_1$)}
    \State Append $S$ to $\mathtt{queryPoints}$, if $S \notin \mathtt{queryPoints}$.
    \If{(response to the query is immediate) or ($t \geq t_1 + T$)}
        \State Call an optimized localization method
        \State Let $(\hat{x},\hat{y})$ be the location obtained from the method;
        \While{($\mathtt{queryPoints} \neq \emptyset $)}
            \State Extract a query sender, say $S'$, from $\mathtt{queryPoints}$;
            \State Send $t_1 $, $t$, $(x_1,y_1)$ and $(\hat{x},\hat{y})$ to $S'$
        \EndWhile
        \State Set $t_1=t$ and $(x_1,y_1)$ = $(\hat{x},\hat{y})$.
    \EndIf
\EndWhile
\end{algorithmic}           \label{algo:main} 
\end{algorithm}
After receiving a message from a sensor, the base station waits until it
gets location information of the sender, $S$. If the processing of the message is immediate, the
base station may send location query to the node $S$. The base station extracts localization points
from  the response obtained from $S$ against the location query and estimates the location of $S$ as
follows:\\
%
\hrule
\begin{algorithmic}[1]
\State Let $(x_1,y_1)$ and $(x_2,y_2)$ be the localized positions of $S$ at $t_1$ and $t_2$
respectively. Let the base station require the position of $S$ at the time $t'$.
\If{($t'\in [t_1,t_2]$)}
    \State Calculate the velocity vector as follows:
    \State ~~~~ $\dot{x} = (x_2 - x_1)/(t_2-t_1)$;  ~~~~~~ $\dot{y} = (y_2 - y_1)/(t_2-t_1)$;
    \State Estimate the position of $s$ at time $t'\in [t_1,t_2$] as follows:
    \State ~~~~ $\hat{x} = x_1 + \dot{x}(t'-t_1)$;  ~~~~~~ $\hat{y} = y_1 + \dot{y}(t'-t_1)$;
\EndIf
\end{algorithmic}
\hrule ~\\[.1mm]
The base station estimates the locations of those sensors only whose events are being processed
recently by the base station. The location of a sensor at a particular time instant on demand are
estimated from the locations obtained in the previous and next localizations nearest to the time
instant.

\begin{figure}[h!]
 \centering
 \includegraphics[width=.9\textwidth,keepaspectratio=true]{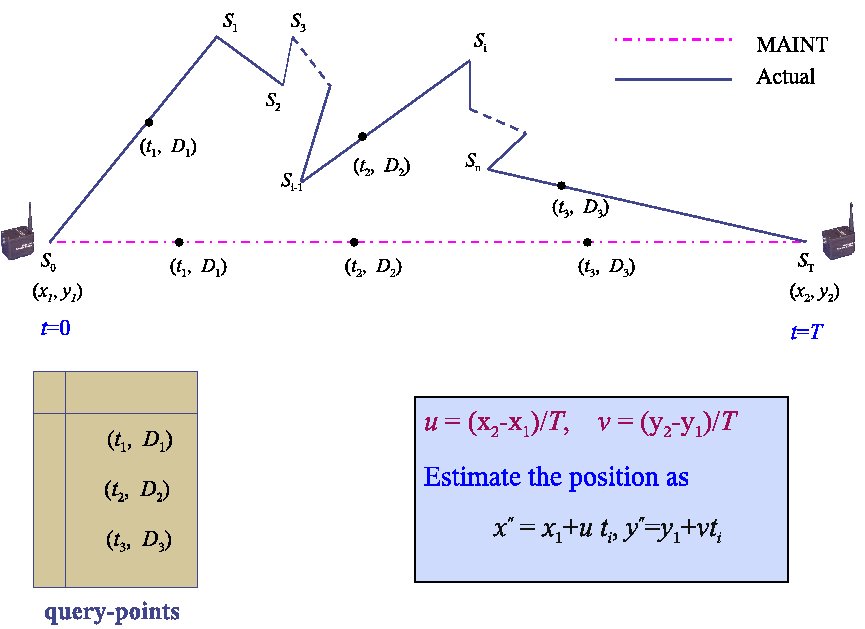}
 \caption{Describing the algorithm with an example.}
 \label{fig:algo}
\end{figure}

We explain the proposed algorithm, MAINT, with an example. Figure~\ref{fig:algo} describes
pictorially the algorithm with an example. Suppose, the sensor calls the localization at time $t_0$
and gets the position $(x_1,y_1)$. Suppose, the sensor receives a query for its location, at time
$t_1$ from the destination $D_1$ (may be a sensor or base station).
 Instead sending the location immediately, the sensor keeps track of the
query inserting $t_1$ and $D_1$ into a list $\mathtt{queryPoints}$. Similarly, it receives the
queries at time $t_2$ and $t_3$ from the destinations $D_2$ and $D_3$ respectively. To keep track
these query points, $t_2$, $D_2$ and $t_3$, $D_3$ are appended in $\mathtt{queryPoints}$. After time
$T$ the MAINT calls the localization to know the actual position $(x_2,y_2)$. The sensor sends the
message consisting of $t_0$, $(x_1,y_1)$ and $T$, $(x_2,y_2)$ corresponding to these two
localization points to the query senders $D_1$, $D_2$ and $D_3$. The query senders find the
locations of the sensor extracting the information from the message. To reduce the message size, the
sensor itself may calculate the velocity from the localization point at $t_0$ and that at $T$. Using
this velocity, calculate the locations at $t_1$, $t_2$ and $t_3$ and sends the locations to $D_1$,
$D_2$ and $D_3$ respectively. It increases arithmetic overhead in the sensor but reduces traffic
through the network.

\section{Energy and Error Analysis} \label{analysis}

Localization in static network is very costly. Finding position of mobile sensors, it needs frequent
localization calls. We assume, the energy consumption is proportional to the number of localization
calls. So we measure energy in terms of number of localization calls. In this work we are reducing
the number of localization calls for the shake of energy saving rather than efficient method.

\subsection{Mobility Model}

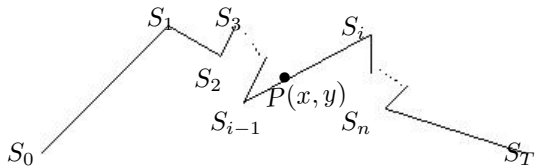
\begin{figure}[!h]
   \centering
    \setlength{\unitlength}{.6mm}
    \begin{picture}(120,37)(0,0)
    \put(12.5,1.5){\includegraphics[width=.5\textwidth,keepaspectratio=true]{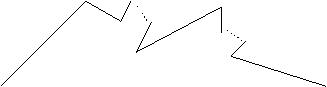}}
    \put(5,0){$S_0$}
    \put(36,30){$S_1$}
    \put(46.5,16.5){$S_2$}
    \put(51,30){$S_3$}
    \put(50,7){$S_{i-1}$}
    \put(79,28){$S_{i}$}
    \put(79,7){$S_{n}$}
    \put(115,0){$S_{T}$}
    \put(65.1,17){\rotatebox{0}{$\bullet$}}
    \put(62,12){\rotatebox{5}{$P(x,y)$}}
    \end{picture}
\caption{A snapshot of Actual Path of Sensor under Random Waypoint Model}
\label{fig:rwp}
\end{figure}

The Random Waypoint (RWP) \cite{JM96,BW02} is a commonly used synthetic model for mobility. We
carried out the simulation study as well as analysis with RWP mobility model. The parameters used in
the model are described as follows:
\begin{itemize}
\item  Each node moves along a zigzag line from one waypoint to the next. The
       next waypoint is selected randomly over a given convex area with two
       parameters \textit{time} and \textit{velocity}.
\item  At the beginning of each leg, a random velocity is drawn from the velocity
       distribution and reach the next waypoint at random time drawn from the
       time distribution.
\end{itemize}
In Figure\,\ref{fig:rwp} the sensor starts from $S_0(x_0,y_0)$ and reaches $S_T(x_T,y_T)$, the point
 at time $T$. The sequence of the waypoints attended by the sensor in the time interval $[0,T]$ is
$S_0,S_1,\cdots,S_n,S_T$.
Let $t_0$ and $t_T$ be  time instances respectively for two consecutive calls of MAINT. The
positions of the sensor are known without any error at the time instances $t_0$ and $t_T$. Estimated
positions of the sensor in between  $t_0$ and $t_T$ may be erroneous.  Without loss of generality,
we assume that $t_0=0$, $t_T=T$ and $(x_0,y_0)=(0,0)$. Because, the  error analysis remains similar
in between any two consecutive calls of MAINT. We assume, at any waypoint $S_{i}$, the sensor draws
the time interval $t_{i+1}$ as well as the velocity vector $(u_{i+1},v_{i+1})$ randomly and
independently. The sensor reaches the next waypoint $S_{i+1}$ after the time $t_{i+1}$ with the
velocity vector $(u_{i+1},v_{i+1})$. For the sake of simplicity we assume, the time interval follows
the exponential distribution with mean $\frac{1}{\lambda}$ and the velocity components $u_{i+1}$ and
$v_{i+1}$ are independent and identically distributed with $Normal(0,\sigma)$ at any waypoint for
$i=0,1,\cdots,n$. Let $P(x,y)$ be the position of of the sensor at a random time $t$ in $(0,T)$, if
the sensor follows the RWP mobility model.

\subsection{Actual Motion Analysis: Related Parameters and Expressions}

From the theory of probability and stochastic process~\cite{M94,B78}, we can say that the event
of occurring waypoints, according to the above mobility model, follows the Poisson Process with
parameter $\lambda$. Consider a random variable $N(t)$ that denotes the number of waypoints in
the interval $(0,t)$. $N(t)$ follows the Poisson distribution with mean $\lambda t$. The probability
mass function (\textit{pmf}) is
\begin{equation}
\Pr(N(t)=k)=\frac{(\lambda t)^k }{k!}e^{-\lambda t}, \mbox{~~~~for~~} k=0,1,2,\cdots,\infty.
\label{eq:PrN}
\end{equation}
The sum $T_i = t_1+t_2+ \cdots + t_i$ (with $T_0\equiv 0$) represents the time occurring $i$th
waypoint, $S_i$. Since $t_i$s are independent and identically distributed following exponential
distribution with parameter $\lambda$ (mean $\frac{1}{\lambda}$), the random variable $T_i$ follows
the distribution $\Gamma(\lambda,i)$. The pdf is
$$
f_{T_i}(\tau)=\frac{\lambda^i \tau^{i-1}e^{-\lambda \tau}}{\Gamma(i)}~~~~where~~0<\tau<\infty.
$$
Let $(X_i,Y_i)$ represent the position of $i$th waypoint, $1\leq i\leq n$.
$X_i$ and $Y_i$ are independent and identically distributed where
\[
  \begin{array}{lclc}
      X_i &=& X_{i-1}+u_it_i= u_1t_1+u_2t_2+\cdots+u_it_i & and \\
      Y_i &=& Y_{i-1}+v_it_i= v_1t_1+v_2t_2+\cdots+v_it_i
  \end{array}
\]
with $(X_0,Y_0)=(0,0)$. The velocity components $u_i$ and $v_i$ are independent and both follow the
distribution $Normal(0,\sigma)$.\\[-2mm]

Given $n$ waypoints have been occurred by time $\tau$.  Let $T_k\mid N(\tau)=n$ denote the waiting
time of the $k$-th waypoint $(1 \leq k \leq n)$ under the given setup.
\begin{result}
\label{res:pdf:Tt_k|n}
 The joint pdf of ~$(T_{k-1},t_k)\mid N(\tau)=n$~ for  ~$2\leq k \leq n$~ is
    \begin{eqnarray*}
    \lefteqn{f_{T_{k-1}t_k}\left((x,y) \mid N(\tau)=n\right)}\\
         &=& \left\{ \begin{array}{ll}
            \frac{n!}{(k-2)!\;(n-k)!}\cdot\frac{x^{k-2}}{\tau^n}\left(\tau-x-y\right)^{n-k},
                                               &    0 < x < \tau, ~ 0 < x +y < \tau\\[2mm]
            0,                                 &    otherwise.
            \end{array}\right.
    \end{eqnarray*}
\end{result}
\proof{
For ~$2\leq k \leq n$, $0 < x < \tau,  0 < x +y < \tau$, the probability element
\[\begin{array}{rcl}
\lefteqn{\Pr((x<T_{k-1} \leq x+dx, ~y<t_k \leq y+dy)\mid (N(\tau)=n))} \\[2mm]
    & = & \frac{\Pr(x<T_{k-1} \leq x+dx,~ y<t_k \leq y+dy~ and~ N(\tau)=n)}{\Pr(N(\tau)=n)}\\[2mm]
    & = & \Pr(x<T_{k-1} \leq x+dx) \cdot \frac{\Pr(y<t_k \leq y+dy\mid T_{k-1}=x)\cdot
                                \Pr((N(\tau)=n)\mid (T_k=x+y))}{\Pr(N(\tau)=n)}  \\[2mm]
    & = & \Pr(x<T_{k-1}\leq x+dx)\cdot \Pr(y<t_k\leq y+dy)
                                 \cdot \frac{\Pr(N(\tau-x-y)=n-k)} {\Pr(N(\tau)=n)}\\[2mm]
    & = & \frac{\frac{\lambda^{k-1}\,x^{k-2}}{\Gamma(k-1)}\,e^{-\lambda x}\,dx\cdot
          \lambda\,e^{-\lambda y}\,dy\cdot \frac{\lambda^{n-k}\,(\tau-x-y)^{n-k}}{(n-k)!}
          \,e^{-\lambda (\tau - x - y)}}{\frac{\lambda^n\,\tau^n}{n!}\,e^{-\lambda \tau}}\\[3mm]
    & = & \frac{n!}{(k-2)!\;(n-k)!}\,\cdot\frac{x^{k-2}}{\tau^n}~(\tau-x-y)^{n-k}\,dx\,dy
\end{array}\]
Hence the pdf.
}

\begin{result}
\label{res:pdf:T_k|n}
  The pdf of ~$T_k\mid N(\tau)=n$~ for  $1\leq k \leq n$  is given by
    \[f_{T_k}(x\mid N(\tau)=n)= \left\{ \begin{array}{lcl}
            \frac{n!}{(k-1)!\;(n-k)!}\cdot\frac{x^{k-1}}{\tau^k}\left(1-\frac{x}{\tau}\right)^{n-k},
                                    &       &   0 < x < \tau\\[2mm]
            0,                      &       &   otherwise.
            \end{array}\right.
    \]
\end{result}
\proof{
From Result~\ref{res:pdf:Tt_k|n}, for $2\leq k \leq n$, the pdf of ($T_{k-1}\mid N(\tau)=n$) is
\[\begin{array}{rcl}
    \lefteqn{f_{T_{k-1}}\left(x\mid N(\tau)=n\right)}\\[1mm]
        & = & \left\{ \begin{array}{lcl}
                    \frac{n!}{(k-2)!\;(n-k)!}\cdot \frac{x^{k-2}}{\tau^n} \cdot
                        \int_0^{\tau - x}\left(\tau-x-y\right)^{n-k}\, dy,
                                            &    &   0 < x < \tau \\[1.5mm]
                    0,                      &    &   otherwise.
                    \end{array}\right.\\[4mm]
        & = & \left\{ \begin{array}{lcl}
                    \frac{n!}{(k-2)!\;(n-k+1)!}\cdot \frac{x^{k-2}}{\tau^n} \cdot
                        \left(\tau-x\right)^{n-k+1},
                                            &    &   0 < x < \tau\\[1.5mm]
                    0,                      &    &   otherwise.
                    \end{array}\right.
\end{array}\]
The pdf of ~$T_k\mid N(\tau)=n$~ for ~$1\leq k \leq n-1$ is
\[\begin{array}{rcl}
    f_{T_k}\left(x\mid N(\tau)=n\right)
        & = & \left\{ \begin{array}{lcl}
                    \frac{n!}{(k-1)!\;(n-k)!}\cdot \frac{x^{k-1}}{\tau^n} \cdot
                        \left(\tau-x\right)^{n-k},
                                            &    &   0 < x < \tau\\[1.5mm]
                    0,                      &    &   otherwise.
                    \end{array}\right.
\end{array}\]
For ~$0 < x < \tau$, the probability element, 
\[\begin{array}{rcl}
 \lefteqn{\Pr(x<T_n \leq x+dx\mid N(\tau)=n)} & \\[3mm]
    & = & \frac{\Pr(x<T_n \leq x+dx~ and~ N(\tau)=n)}{\Pr(N(\tau)=n)}  \\[3mm]
    & = & \frac{\Pr(x<T_n \leq x+dx)\cdot \Pr(N(\tau)=n\mid T_n=x)}{\Pr(N(\tau)=n)}  \\[3mm]
    & = & \frac{\Pr(x<T_n \leq x+dx)\cdot \Pr(N(\tau-x)=0)}
                                            {\Pr(N(\tau)=n)}\\[3mm]
    & = & \frac{\lambda^n\,x^{n-1}}{\Gamma(n)}\,e^{-\lambda x}\,dx\cdot e^{-\lambda (\tau-x)}/
\left(\frac{\lambda^n\,\tau^n}{n!}\,e^{-\lambda \tau}\right)\\[3mm]
    & = & n\cdot\frac{x^{n-1}}{\tau^n}\,dx
\end{array}\]
Hence, the pdf of ~$T_k\mid N(\tau)=n$~ is followed, for  $1\leq k \leq n$.
}

\begin{result}
    \label{res:Exp:T_k|n}
    $E(T_k\mid N(\tau)=n) = \frac{k\,\tau}{n+1}$ and
    $E(T_k^2\mid N(\tau)=n) = \frac{k(k+1)\,\tau^2}{(n+1)(n+2)}$, for~ $1\leq k \leq n$.
\end{result}
\proof{
Using the pdf as in Result~\ref{res:pdf:T_k|n}, ~for ~$1\leq k \leq n$, we may write the expected
values $E(T_k^m\mid N(\tau)=n)$ as follows:
\[\begin{array}{rcl}
    {E(T_k^m\mid N(\tau)=n)}
    & = & \int_0^{\tau}x^m\,f_{T_k}\left(x\mid N(\tau)=n\right)\, dx\\[2mm]
    & = & \frac{n!\;\tau^m}{(k-1)!\,(n-k)!} \int_0^{\tau}\left(\frac{x}{\tau}\right)^{m+k-1}
                    \,\left(1-\frac{x}{\tau}\right)^{n-k}\, \frac{dx}{\tau}\\[3mm]
    & = & \frac{n!\;\tau^m}{(k-1)!\,(n-k)!} \int_0^{1}x^{m+k-1}\,\left(1-x\right)^{n-k}\,dx\\[3mm]
    & = &
    \frac{n!\;\tau^m}{(k-1)!\,(n-k)!}\cdot\frac{\Gamma(m+k)\;\Gamma(n-k+1)}{\Gamma(n+m+1)}\\[3mm]
    & = & \frac{n!\;(m+k-1)!}{(k-1)!\;(n+m)!} ~ \tau^m
\end{array}\]
Substituting, $m$ by $1$ and $2$,~  we may have the expectations ~
$E(T_k\mid N(\tau)=n)$ ~ and ~ $E(T_k^2\mid N(\tau)=n)$, ~ for~ $1\leq k \leq n$. Hence the result
follows.
}

Let $t_k\mid N(\tau)=n$ represent the time interval between the $(k-1)$th and the $k$th waypoints
under the given setup when we are given that exactly $n$ waypoints have occurred in the interval
$(0,\tau)$.
\begin{result}
    \label{res:pdf:t_k|n}
    The density function of ~$t_k\mid N(\tau)=n$,~ for ~$1\leq k \leq n$~ is
    \[
        f_{t_k}(y \mid N(\tau)=n) =  \left\{\begin{array}{ll}
                        \frac{n}{\tau}\left(1-\frac{y}{\tau}\right)^{n-1},  & 0\leq y<\tau\\[2mm]
                        0,                                                  & otherwise.
    \end{array}\right.\]
\end{result}
\proof{
The random variable $T_1 = t_1$. The distribution of ~($t_1\mid N(\tau)=n$) is same
as the distribution of  ($T_1\mid N(\tau)=n$). From the Result~\ref{res:pdf:Tt_k|n}, the
distribution of ~ $t_k\mid N(\tau)=n$ ~for  $2\leq k \leq n$ is
\[\begin{array}{rcl}
    \lefteqn{f_{t_k}\left(y\mid N(\tau)=n\right)}\\[1mm]
        & = & \left\{\begin{array}{ll}
                    \frac{n!}{(k-2)!(n-k)!}\cdot \frac{(\tau-y)^{n-1}}{\tau^n}
                    \int_0^{\tau-y}(\frac{x}{\tau-y})^{k-2}
                        (1-\frac{x}{\tau-y})^{n-k}\,\frac{dx}{\tau-y},
                                            &    0\leq y < \tau\\[2mm]
                    0,                      &    otherwise.
                    \end{array}\right.\\[4mm]
        & = & \left\{ \begin{array}{lcl}
                    \frac{n!}{(k-2)!(n-k)!}\cdot \frac{(\tau-y)^{n-1}}{\tau^n}
                    \int_0^{1}t^{k-2}\,(1-t)^{n-k}\, dt,
                                            &    &   0\leq y < \tau\\[2mm]
                    0,                      &    &   otherwise.
                    \end{array}\right.\\[4mm]
        & = & \left\{ \begin{array}{lcl}
                    \frac{n!}{(k-2)!(n-k)!}\cdot \frac{(\tau-y)^{n-1}}{\tau^n}
                    \cdot \frac{\Gamma(k-1)\,\Gamma(n-k+1)}{\Gamma(n)},
                                            &    &   0\leq y < \tau\\[2mm]
                    0,                      &    &   otherwise.
                    \end{array}\right.
\end{array}\]
Thus the pdf of ~$t_k\mid N(\tau)=n$ ~for $1\leq k \leq n$ is
\[\begin{array}{rcl}
    f_{t_k}\left(y\mid N(\tau)=n\right)
        & = & \left\{ \begin{array}{lcl}
                    \frac{n}{\tau}\,\left(1-\frac{y}{\tau}\right)^{n-1},
                                            &    &   0\leq y < \tau\\[3mm]
                    0,                      &    &   otherwise.
                    \end{array}\right.\\[-3mm]
\end{array}\]
}

\begin{result}
 \label{res:Exp:t_k|n}
 $E(t_k\mid N(\tau)=n)$ = $\frac{\tau}{n+1}$ and
 $E(t_k^2\mid N(\tau)=n)$ = $\frac{2\tau^2}{(n+1)(n+2)}$
\end{result}
\proof{
Using the pdf as in Result~\ref{res:pdf:T_k|n}, ~ for ~ $1\leq k \leq n$, ~ we have
\[\begin{array}{rcl}
E(t_k^m\mid N(\tau)=n)  & = & \int_0^{\tau}y^m\,f_{t_k}\left(y\mid N(\tau)=n\right)\,dy\\[3mm]
                        & = & n\,\tau^m \int_0^{\tau}\left(\frac{y}{\tau}\right)^m
                            \,\left(1-\frac{y}{\tau}\right)^{n-1}\, \frac{dy}{\tau}\\[3mm]
                        & = & n\,\tau^m \int_0^1 y^m\,\left(1-y\right)^{n-1}\,dy\\[3mm]
                        & = & n\,\tau^m\cdot\frac{\Gamma(m+1)\;\Gamma(n)}{\Gamma(n+m+1)}\\[3mm]
                        & = & \frac{n!\;m!}{(n+m)!} ~ \tau^m
\end{array}\]
Putting $m =1$ and $m=2$ in the above relation,  we can have  $E(t_k\mid N(\tau)=n)$ and
$E(t_k^2\mid N(\tau)=n)$, for $1\leq k \leq n$.
}

\begin{result}
\label{res:Exp:X_k|n}
 $E(X_k\mid N(\tau)=n) = 0$ and $E(X_k^2\mid N(\tau)=n)$ =
$\frac{2k\,\sigma^2\,\tau^2}{(n+1)\,(n+2)}$,  ~for~ $1 \leq k \leq n$.
\end{result}
\proof{
 $X_k = \sum_{i=1}^k u_i\,t_i $, for $1 \leq k \leq n$~ and $t_k$\,s and $u_k$\,s are independent.
\[\begin{array}{rcl}
    E(X_k\mid N(\tau)=n)
    & = & E\left(\left(\sum_{i=1}^k u_i\,t_i\right)\mid N(\tau)=n\right)\\[2mm]
    & = & \sum_{i=1}^kE\left(u_i\right)\, E\left(t_i\mid N(\tau)=n\right),\\
    &   &   \hspace*{.5cm} \mbox{since, $u_i$\,s are independent on $t_i$\,s and $N(\tau)=n$}\\
    & = & 0,\hspace*{2.6cm} \mbox{since $E(u_i)=0$.}
\end{array}\]
Similarly,  ~ $E(X_k^2\mid N(\tau)=n)$
\[\begin{array}{rl}
    & =  E\left(\left(\sum_{i=1}^k u_i\,t_i\right)^2\mid N(\tau)=n\right)\\[3mm]
    & =  E\left(\left(\sum_{i=1}^k u_i^2\,t_i^2 + 2\, \sum_{i<j=1}^k
          u_i\,u_j\,t_i\,t_j\right) \mid N(\tau)=n\right)\\[1mm]
    & =  \sum\limits_{i=1}^k E(u_i^2)\,E(t_i^2\mid N(\tau)=n) +
          2\sum\limits_{i<j=1}^k E(u_i)\,E(u_j)\,E(t_i\,t_j \mid N(\tau)=n)\\[.1mm]
    & =  \sigma^2\sum\limits_{i=1}^k E(t_i^2\mid N(\tau)=n),
          ~~~~~~~~~~ \mbox{since, $E(u_i)=0$ and $E(u_i^2)=\sigma^2$}\\[3mm]
    & =  \frac{2k\,\sigma^2\,\tau^2}{(n+1)\,(n+2)},
          ~~~~~~~~~~~~~~~~~~~~~ \mbox{using Result~\ref{res:Exp:t_k|n}.}\\[-3mm]
\end{array}\]
}

\subsubsection{Actual Position of Sensor}

We analyze the motion of the sensor in between two consecutive calls of localization. Because,
the pattern of the motion remains similar in between any two consecutive localization points.
Let $P(x,y)$ be the position of the sensor at a random time $t$, $0< t < T$. Consider
the random variable $(X,Y)$ that represents the position of $P$. Let $i$ waypoints occur
in the interval $(0,t)$, i.e. $0\leq T_i\leq t < T_{i+1}$. Given $N(t)=i$. Then we have
\begin{center}
  $X = X_i + (t-T_i)u_{i+1}$ ~~~~~~~~ $Y = Y_i + (t-T_i)v_{i+1}$
\end{center}
for ~ $i=0,1,2,\ldots,\infty$ where $T_0 = 0$ and $(X_0,Y_0)=(0,0)$.

\begin{theorem}
 \label{cor:Exp:X|i}
 $E(X\mid N(t)=i) = 0$ and $E(X^2\mid N(t)=i)$ = $\frac{2}{i+2}\,\sigma^2\,t^2$.
\end{theorem}
\proof{
Consider ~$X = X_i + (t-T_i)\,u_{i+1}$, given ~$N(t) = i$, for a fixed ~$t\in (0,T)$.
\[\begin{array}{rcl}
    E(X\mid N(t)=i) & = & E(X_i\mid N(t)=i) + E((t-T_i)\mid N(t)=i)\,E(u_{i+1})\\
                    & = & 0, ~~~~~~~~~~~~~~
                                 \mbox{using ~$E(u_{i+1})=0$~ and ~Result~\ref{res:Exp:X_k|n}.}
\end{array}\]
Similarly, ~$E(X^2\mid N(t)=i)$
\[\begin{array}{rcl}
        & = & E(X_i^2\mid N(t)=i) + E(u_{i+1}^2)\,E((t-T_i)^2\mid N(t)=i)\\[.1mm]
        &   & \hspace*{5cm}   + 2\,E(u_{i+1})\,E((t-T_i)\mid N(t)=i)\\
        & = & E(X_i^2\mid N(t)=i) + \sigma^2\,(t^2 - 2t\;E(T_i\mid N(t)=i) + E(T_i^2\mid
                                                                        N(t)=i))\\[.5mm]
        & = & \frac{2i\,\sigma^2\,t^2}{(i+1)\,(i+2)} + \sigma^2\,\left(t^2 - 2t\;\frac{i\,t}{i+1}
               + \frac{i\,(i+1)\,t^2}{(i+1)\,(i+2)}\right), ~~~~
                     \mbox{using Result~\ref{res:Exp:T_k|n} and \ref{res:Exp:X_k|n}.}\\
        & = & \frac{2}{i+2}\;\sigma^2\,t^2. \\[-3mm]
\end{array}\]
}

\begin{theorem}               \label{theo:Exp:X}
 The expectation of $X$ and $X^2$ are given by $E(X)=0$ and
    $$E(X^2)=2\sigma^2\left(\frac{t}{\lambda} - \frac{1}{\lambda^2}+\frac{1}{\lambda^2}
                        e^{-\lambda t}\right).$$
\end{theorem}
\proof{
The random variable $X$ represents the $x$-coordinate of the sensor at time $t$. In the a
particular time instant the expected value of $X^2$ is given by
\[\begin{array}{rcl}
 E(X^2) & = & E[E(X^2\mid N(t)=i)]\\[2mm]
        & = & E\left(\frac{2\sigma^2t^2}{i+2}\right), ~~\mbox{using
                                                    Theorem~\ref{cor:Exp:X|i}}\\[3mm]
        & = & 2\sigma^2t^2\sum\limits_{i=0}^{\infty}\frac{1}{i+2}\Pr(N(t)=i)\\[4mm]
        & = & 2\sigma^2t^2\sum\limits_{i=0}^{\infty}\frac{1}{i+2}
                                            \,\frac{(\lambda t)^i}{i\,!}e^{-\lambda t}\\[5mm]
        & = & 2\sigma^2t^2\left(\frac{1}{\lambda t} - \frac{1}{\lambda^2 t^2}
                                            +\frac{1}{\lambda^2 t^2} e^{-\lambda t}\right)\\[4mm]
        & = & 2\sigma^2\left(\frac{t}{\lambda} - \frac{1}{\lambda^2}
                                    +\frac{1}{\lambda^2} e^{-\lambda t}\right)\\[-2mm]
\end{array}\]
}

\subsection{Estimation by MAINT and Error Analysis}

\begin{figure}[!h]
    \setlength{\unitlength}{.7mm}
    \begin{picture}(418,57)(-30,0)
    \put(11.5,1.2){\includegraphics[width=.6\textwidth,keepaspectratio=true]{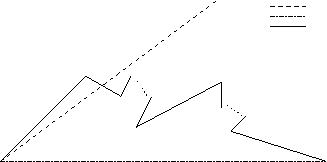} }
    \put(110,51.5){\tiny \it MADRD}
    \put(110,48.5){\tiny \it MAINT}
    \put(110,45.5){\tiny \it ACTUAL }
    \put(5,0){$S_0$}
    \put(36,29){$S_1$}
    \put(47,17){$S_2$}
    \put(51,29){$S_3$}
    \put(50,9){$S_{i-1}$}
    \put(81,28){$S_{i}$}
    \put(77,8.5){$S_{n}$}
    \put(116,0){$S_{T}$}
    \put(81,53){$S_{T}'$}
    \put(65.2,41){\rotatebox{5}{$\bullet$}}
    \put(65,37){\rotatebox{0}{$P''$}}
    \put(61,15){\rotatebox{0}{$\bullet$}}
    \put(61,11){\rotatebox{5}{$P$}}
    \put(70,0){$\bullet$}
    \put(68,3){\rotatebox{5}{$P'$}}
    \end{picture}
\caption{Showing position estimates at an intermediate point by MADRD and MAINT.}
\label{ErrEnrg}
\end{figure}
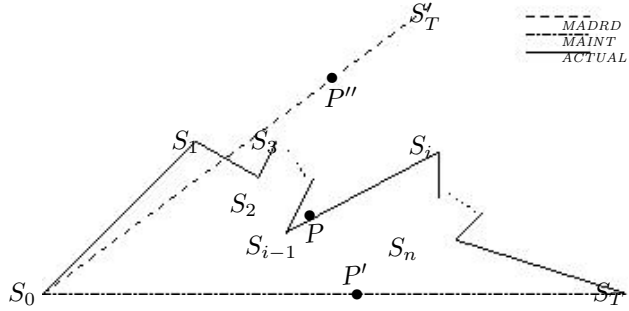

Assume two consecutive calls of MAINT occur at the times $0$ and $T$. In Figure\,\ref{ErrEnrg},
$S_0,S_1,\ldots,S_n,S_T$ is the actual path of the sensor in between the times $0$ and $T$.
Let $P(x,y)$ be the actual position of the sensor at a random time $t\in (0,T)$ when it follows the
said RWP mobility model. Let $P'\,(x_{est},y_{est})$ be the estimated position of the sensor at $t$
according to MAINT. Let  $(X_{est},Y_{est})$ denotes the random variable to estimated position
$P'$ by MAINT at time $t$. Then we have
                 $$X_{est}=\frac{X_T}{T}t,~~~~Y_{est}=\frac{Y_T}{T}t$$
where $(X_T,Y_T)$ is the random position of the sensor at time $T$.

\subsubsection{Error Analysis}
In the analysis, we consider RWP mobility model. We assume the waypoints follows the Poisson
process. Since localizations occur at time $0$ and $T$, we consider the motion in the time interval
$[0,T]$. For error calculation we take a location estimation at a random time $t \in [0,T]$. Due
to the memory less property of Poison process, we may break the complete scenario in two independent
Poisson processes with same parameter, one in the interval $[0,t]$ and another in $[t,T]$. As a
whole, these two processes represent the same process as in $[0,T]$. Suppose, $N'(\tau)=m$ denote
the event that $m$ waypoints occur in $[t,T]$.
\begin{theorem}         \label{theo:N'}
$\Pr(N'(\tau)=m) = \Pr(N(\tau-t)=m)$.
\end{theorem}
\proof{
It follows from the memory less property of Poisson process~\cite{B78}. 
}

\begin{theorem}             \label{theo:NN'}
$\Pr(N(t)=i,N'(T)=j) = \frac{\lambda ^{i+j}e^{-\lambda T}}{i!\,j!}t^i (T-t)^{j}$.
\end{theorem}
\proof{
From the Poisson process we can say that the events $N(t)=i$ and $N'(T)=j$ are independent.
Therefore,
            $$\Pr(N(t)=i,N'(T)=j) = \Pr(N(t)=i)\Pr(N'(T)=j).$$
From the equation~(\ref{eq:PrN}) and the Theorem~\ref{theo:N'} we have
\begin{eqnarray*}
\Pr(N(t)=i,N'(T)=j) &=& \Pr(N(t)=i)\Pr(N(T-t)=j) \\
                    &=& \frac{(\lambda t)^i}{i!}\,e^{-\lambda t}\cdot\frac{(\lambda(T-t))^{j}}{j!}\,
                        e^{-\lambda(T-t)}              \\
                    &=& \frac{\lambda ^{i+j}\,e^{-\lambda T}}{i!\,j!}t^i (T-t)^{j}
\end{eqnarray*}
}

\paragraph{Expected Error in MAINT}

Consider a random time $t$ in the interval $(0,T)$. Let $S_1$, $S_2$, $\cdots$, $S_i$ be $i$
waypoints occurred in $[0,t]$ and $j$ waypoints $S_{i+1}$, $S_{i+2}$, $\cdots$, $S_{i+j}$
occur in the time interval $[t,T]$.  Under this setup, the actual position of the sensor at time $t$
is given by the random variable $(X,Y)$ where
               $$X=X_i+(t-T_i)u_{i+1}$$        $$ Y=Y_i+(t-T_i)v_{i+1}.$$

Since, the process in $[0,t]$ and in $[t,T]$ are independent to each other, we may assume
the waypoints $S_{i+1}$, $S_{i+2}$, $\cdots$, $S_{i+j}$ occurs just like the system starts from the
time $t$ where the position of the sensor is $(X,Y)$. Due to the memory less property of the
Poisson Process, we may obtain the time occurrences of the waypoints $S_{i+1}$, $S_{i+2}$, $\cdots$,
$S_{i+j}$ form the same Poisson process over the time interval $[0,T-t]$ with an additional time
$t$. Let $T'_k$ and $t'_k$ denote the time of occurrence of the position of the waypoint $S_{i+k}$
and time interval between two waypoints of the motion of the sensor in $[0,T-t]$. If $(X'_k,Y'_k)$
denote the the random position of the waypoint $S_{i+k}$ taking $(X,Y)$ as the origin, we have
    $$X'_k = X'_{k-1} + t'_ku_{i+k} = \sum_{m=1}^k t'_mu_{i+m},$$
    $$Y'_k = Y'_{k-1} + t'_kv_{i+k} = \sum_{m=1}^k t'_mv_{i+m}$$
for $k= 0,1,2, \ldots, \infty$ where $T'_0=0$ and $(X'_0,y'_0)=(0,0)$.

The random velocity vector at any waypoint is independent to time of occurrence of the waypoint. So
$(X_i,Y_i)$ and $(X'_k,Y'_k)$ are independent for $i\geq 1, k\geq 1$. If we assume the coordinates
of the waypoint $S_{i+j}$ in the whole process over $[0,T]$ as $(X_{i+j},Y_{i+j})$, we have
$X_{i+j}=X+X'_j$, $Y_{i+j}=Y+Y'_j$ and $T_{i+j}=t+T'_j$. If $(X',Y')$ denote the position of the
sensor at time $T$ due to the process over $[0,T-t]$ under the condition that $N'(T)=j$, i.e.,
$N(T-t) = j$ then
\begin{center}
         $ X' = X'_j+(T-t-T'_j)u_{i+j+1}, ~~~~~~~~     Y' = Y'_j+(T-t-T'_j)v_{i+j+1}.$
\end{center}
The position of the sensor at time $T$, $(X_T,Y_T)$, due to the process over $[0,T]$ under
the conditions that $N(t)=i$ and $N'(T)=j$, may be obtained as
\begin{center}
  $ X_T = X+X'_j+(T-t-T'_j)u_{i+j+1} = X + X',$ \\[2mm]
  $Y_T = Y+Y'_j+(T-t-T'_j)v_{i+j+1} = Y + Y'.$
\end{center}
Therefore, the estimated positions at time $t$ may written as:
\begin{center}
         $ X_{est} = \frac{X+X'}{T}\,t, ~~~~~~~~~~ Y_{est} = \frac{Y+Y'}{T}\,t$
\end{center}
Let $error_t$ denote the expected squared error in the position estimation by MAINT at a random
time $t$ in $[0,T]$. Thus, $error_t$ can be expressed as:
\begin{equation}
\begin{array}{rcl}
 error_t & = & E\left[(X-X_{est})^2+(Y-Y_{est})^2\right] \\[1mm]
         & = & 2E\left[\left(X-X_{est}\right)^2\right], ~~ \mbox{since $X-X_{est}$ and $Y-Y_{est}$
                                                                are iid} \\[2mm]
         & = & 2E\left[\left(X-\frac{X+X'}{T}\,t\right)^2\right]  \\[3mm]
         & = & 2E\left[\left\{\left(1-\frac{t}{T}\right)X-\frac{t}{T}X'\right\}^2\right] \\[2mm]
         & = & 2\left[\left(1-\frac{t}{T}\right)^2 E(X^2)+\frac{t^2}{T^2}E\left(X'^2\right)
                            - 2\frac{t}{T}\left(1-\frac{t}{T}\right)E(XX')\right]
\end{array}                                                                     \label{eq:errot_t}
\end{equation}

\begin{theorem}  \label{theo:Exp:X'}
 The expectation of $X'$ and $X'^2$ are given by $E(X')=0$ and
\begin{center}
$E(X'^2)=2\sigma^2\left\{\frac{T-t}{\lambda} - \frac{1}{\lambda^2}
             +\frac{1}{\lambda^2} \,e^{-\lambda (T-t)}\right\}. $
\end{center}
\end{theorem}
\proof{
We have seen that $X' = X'_j+(T-t-T'_j)u_{i+j+1}$ is the $x$-coordinate of the
sensor at time $T-t$ in the process over $[0,T-t]$. So $X'$ has similar properties as $X$ except
$N'(T)=j$, i.e. $N(T-t)=j$ instead of $N(t)=i$. The event $N'(T)$ is independent with the
scenario prior to the time $t$, i.e. independent with $N(T)$. Thus, the result
follows from Theorem~\ref{theo:Exp:X} replacing $T-t$ instead of $t$.
}

\begin{theorem}\label{theo:Exp:XX'|ij}
Given  $N(t) = i$ and $N'(T)=j$ for a particular time $t\in [0,T]$. The expectation of $XX'$ may be
given as
\begin{center}
    $E\left[XX'\mid N(t)=i,N'(T)=j\right]=\frac{\sigma^2t(T-t)}{(i+1)(j+1)},
                                ~~~~~~~~ for ~~ i,j\geq 0.$
\end{center}
\end{theorem}
\proof{
Under the given condition $N(t) = i$ and $N'(T)=j$ for a particular time $t\in [0,T]$, we know
$X=X_i+(t-T_i)u_{i+1}$ and $X' = X'_j+(T-t-T'_j)u_{i+j+1}$ as stated earlier. Therefore, we have
\[\begin{array}{rcl}
 \lefteqn{E\left[XX'\mid N(t)=i,N'(T)=j\right]} \\[1mm]
    & = &   E\left[\left\{X_i + (t-T_i) u_{i+1}\right\}
                        \left\{X'_j + (T-t-T'_j) u_{i+j+1}\right\}\right] \\[1mm]
    & = &   E(t-T_i)\left\{E\left(u_{i+1}X'_j\right) + E\left(T-t-T'_j\right)
                        E\left(u_{i+1}u_{i+j+1}\right)\right\} \\[1mm]
    & = &   \left\{\begin{array}{lrl}
            E(t-T_i)E\left(u_{i+1}X'_j\right),      & ~~~~~~ for & i\geq 0, ~ j \geq 1 \\[1mm]
            (T-t)E(t-T_i)E\left(u_{i+1}^2\right),   &        for & i\geq 0, ~ j = 0
            \end{array}\right.\\[4mm]
    & = &   \left\{\begin{array}{lrl}
            E(t-T_i)E\left(t'_1\right)E\left(u_{i+1}^2\right),  & ~~~~~~ for & i\geq 0, ~ j\geq 1
                                                                                \\[1mm]
            (T-t)E(t-T_i)E\left(u_{i+1}^2\right),               &        for & i\geq 0, ~ j = 0
            \end{array}\right.\\[4mm]
    & = &   \left\{\begin{array}{lrl}
            \sigma^2 \frac{T-t}{j+1}(t-\frac{it}{i+1}),  & ~~~~~~ for & i\geq 0, ~ j\geq 1 \\[1mm]
            \sigma^2 (T-t)(t-\frac{it}{i+1}),            &        for & i\geq 0, ~ j= 0
            \end{array}\right.\\[5mm]
    & = &   \frac{\sigma^2\, t(T-t)}{(i+1)(j+1)}, ~~~~~~ for ~~ i\geq 0, ~ j\geq 0\\[-5mm]
\end{array}\]
}

\begin{theorem}            \label{theo:Exp:XX'}
    For a particular $t\in [0,T]$, the expectation of $XX'$ is given as
\begin{center}
    $E(XX')=\frac{\sigma^2}{\lambda^2}\left\{1-e^{-\lambda t}
             -e^{-\lambda (T-t)} +e^{-\lambda T}\right\}.$
\end{center}
\end{theorem}
\proof{
The random variables $X$ and $X'$ represent the positions of the sensor in the decomposed motions
of the sensor into intervals $[0,t]$ and $[t,T]$ as discussed earlier. The expectation of $XX'$ is
given as
\[\begin{array}{rcl}
 E(XX')  & = &  E\left[E\left[XX'\mid N(t)=i,N'(T)=j\right]\right] \\[2mm]
         & = &  E\left[\frac{\sigma^2t(T-t)}{(i+1)(j+1)}\right],
                                 ~~~~ \mbox{using Theorem~\ref{theo:Exp:XX'|ij}}\\[2mm]
         & = & 
\sum\limits_{i=0}^{\infty}\sum\limits_{j=0}^{\infty}\left[\frac{\sigma^2t(T-t)}{(i+1)(j+1)}\,
                                            \cdot \Pr(N(t)=i,N'(T)=j)\right]\\[3mm]
         & = & 
\sum\limits_{i=0}^{\infty}\sum\limits_{j=0}^{\infty}\left[\frac{\sigma^2t(T-t)}{(i+1)(j+1)}\cdot
                \frac{\lambda^{i+j}\;t^i(T-t)^j}{i!\;j!}\,e^{-\lambda T}
                \right],                         \mbox{by Theorem\,\ref{theo:NN'}}\\[3mm]
         & = &  \frac{\sigma^2}{\lambda^2}\,e^{-\lambda T}
                \sum\limits_{i=0}^{\infty}\frac{\lambda^{i+1}t^{i+1}}{(i+1)!}
                \sum\limits_{j=0}^{\infty} \frac{\lambda^{j+1}(T-t)^{j+1}}{(j+1)!}\\[2mm]
         & = &  \frac{\sigma^2}{\lambda^2}\,e^{-\lambda T}
                \sum\limits_{i=1}^{\infty}\frac{\lambda^{i}t^{i}}{i!}
                \sum\limits_{j=1}^{\infty} \frac{\lambda^{j}(T-t)^{j}}{j!}\\[3mm]
         & = &  \frac{\sigma^2}{\lambda^2}\,e^{-\lambda T}\left(e^{\lambda t}-1\right)
                \left(e^{\lambda (T-t)}-1\right)\\[3mm]
         & = &  \frac{\sigma^2}{\lambda^2}\left\{1-e^{-\lambda t} -e^{-\lambda (T-t)}
                +e^{-\lambda T}\right\}\\[-4mm]
\end{array}\]
}

Using the Theorems~\ref{theo:Exp:X},\ref{theo:Exp:X'},\ref{theo:Exp:XX'}, the
equation~(\ref{eq:errot_t}) reduces to the following:
\[\begin{array}{rcl}
 error_t & = &  2\left[\left(1-\frac{t}{T}\right)^2 E(X^2)+\frac{t^2}{T^2}E\left(X'^2\right)
                - 2\frac{t}{T}\left(1-\frac{t}{T}\right)E(XX')\right] \\[3mm]
         & = &  \frac{4\sigma^2}{\lambda^2 T^2}\left[
                t^2 \left\{\lambda(T-t) - 1 + e^{-\lambda(T-t)}\right\} +
                (T-t)^2\left(\lambda t - 1 + e^{-\lambda t}\right)\right. \\[2mm]
         &   &  \left. - \left(1+e^{-\lambda T}\right) t(T-t) + t(T-t)\,e^{-\lambda t}
                + t(T-t)\,e^{-\lambda (T - t)}\right]
\end{array}\]

Therefore, the average of squared error, denoted by $error(avg)$, in the location estimation by
MAINT is given as follows:
\[\begin{array}{rcl}
error(avg)  & = &   \frac{1}{T}\int\limits_0^T error_t~dt                 \\[2mm]
            & = &   \frac{4\sigma^2}{\lambda^2 T^3}\left[
                    \int_0^T t^2 \left\{\lambda(T-t) - 1 + e^{-\lambda(T-t)}\right\}\,dt\right.
                                                                           \\[2mm]
            &   &   + \int_0^T(T-t)^2\left(\lambda t-1+e^{-\lambda t}\right)\,dt
                    - \left(1+e^{-\lambda T}\right)\int_0^T t(T-t)\,dt\\[2mm]
            &   &   + \left. \int_0^T  t(T-t)\,e^{-\lambda t}\,dt
                    + \int_0^T  t(T-t)\,e^{-\lambda (T - t)}\,dt \right]  \\[2mm]
            & = &   \frac{4\sigma^2}{\lambda^2 T^3}\left[
                    2 \int_0^T t^2 \left\{\lambda(T-t) - 1 + e^{-\lambda(T-t)}\right\}\,dt
                                                                      \right.  \\[2mm]
            &   &   - \left(1+e^{-\lambda T}\right)\frac{T^3}{6}
                    + \left. 2\int_0^T  t(T-t)\,e^{-\lambda (T - t)}\,dt \right],  \\
            &   &   \hspace*{2cm} \mbox{by fundamental theory of integral calculus}\\[2mm]
            & = &   \frac{4\sigma^2}{\lambda^2 T^3}\left[
                    2 \int_0^T \left(\lambda T\, t^2 - \lambda t^3 - t^2\right)\,dt
                    - \left(1+e^{-\lambda T}\right)\frac{T^3}{6} \right.   \\[1mm]
            &   &   + \left. 2T\,e^{-\lambda T}\int_0^T t\,e^{\lambda t}\,dt\right]\\[2mm]
            & = &   \frac{4\sigma^2}{\lambda^2 T^3}\left[
                    2\left(\frac{\lambda T^4}{12} - \frac{T^3}{3}\right)
                    - \left(1+e^{-\lambda T}\right)\frac{T^3}{6} \right.   \\[2mm]
            &   &   + \left. 2T\,e^{-\lambda T}\left(\frac{1}{\lambda^2} -
                      \frac{1}{\lambda^2}e^{\lambda T}
                      + \frac{T}{\lambda}e^{\lambda T}\right)\right]  \\[2mm]
            & = &   \frac{4\sigma^2}{\lambda^2 T^3}
                    \left(\frac{\lambda T^4}{6} - \frac{5T^3}{6} + \frac{2T^2}{\lambda}
                     - \frac{2T}{\lambda^2} + \frac{2T}{\lambda^2}\,e^{-\lambda T}
                      - \frac{T^3}{6}\,e^{-\lambda T} \right)         \\[2mm]
            & = &   \frac{2\sigma^2}{3\lambda^2}\left[\lambda T - 5  + \frac{12}{\lambda T}
                    - \frac{12}{\lambda^2 T^2} + \frac{12}{\lambda^2 T^2} \, e^{-\lambda T}
                    - e^{-\lambda T} \right]
\end{array}\]
If we assume $T$ and $\lambda$ grow with $\frac{T}{\lambda} = constant = C$, we get
\[\begin{array}{rcl}
    \lefteqn{error(avg)}\\[2mm]
            & = &   \frac{2\sigma^2}{3}\left[C - \frac{5\,C^2}{T^2}  + \frac{12\,C^3}{T^4}
                    - \frac{12\,C^4}{T^6} + \frac{12\,C^4}{T^6} \, e^{-T^2/C}
                    - e^{-T^2/C} \right] \\[3mm]
            & = &   \frac{2\sigma^2\, C}{3}, ~~~~~as~~~ T\rightarrow\infty ~~with~~ T=C\,\lambda
\end{array}\]

From the above result, we see that the average error approaches to zero when $T$, the time period
of localizations, tends to zero. The error grows as $T$ becomes large. It is very important to see
that as $\lambda$ becomes very large, i.e., sensor changes its direction more frequently, the error
becomes very small. If both $T$ and $\lambda$ grows with constant ratio i.e., $\frac{T}{\lambda} =
constant$, the error approaches to a constant value. Therefore, if we have prior knowledge of the
rate of direction of the motion sensors, we well control the energy with an acceptable level of
error by adjusting the value of $C$.

\section{Analysis by Simulations} \label{sim:res}

Simulation studies were carried out using ns-2~\cite{NS} to compare the performance of the proposed
technique with that of MADRD. In the simulation study, we concentrated mainly on the average error
distance for different number of localization counts. We assume that the sensors move with RWP
mobility model with parameters as in Table~\ref{Sim:Param}.
\begin{table}[h] \vspace*{-2mm}
\centering
\begin{tabular}{|l|l|}
\hline
Mobility model                              & Random Waypoint Model\\ \hline
Velocity components distribution            & $Normal(0,\sigma)$, $\sigma = 5.0$ unit\\ \hline
Time gap between waypoints distribution     & Exponential mean $\frac{1}{\lambda} = 10.0$ sec\\
 \hline
\end{tabular}
\caption{Relevant parameters used in simulation } \label{Sim:Param}
\end{table} \vspace*{-1mm}
In this work, velocity components are chosen from independent Normal distribution and time interval
between any pair of consecutive waypoints is chosen from the Exponential distribution. During
the simulation, we use the parameters described in Table~\ref{Sim:Param}. In this model, the time
is measured in $sec$. The velocities are measured in $unit/sec$. The error in position are
measured in distance unit (i.e., $unit$ as in Table~\ref{Sim:Param}).

\begin{figure}[h] \vspace*{-3mm}
    \centerline{\includegraphics[width=.7\textwidth]{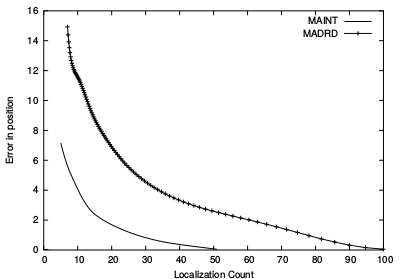} }
\caption{Average error and localization counts for MADRD and the proposed technique.}
\label{ErrEnrgFig} \vspace*{-2.5mm}
\end{figure}

The simulation process was carried out over for a time span of $100\,secs$. Using MADRD and MAINT,
we estimate the position of a mobile sensor at a random time in $[0,100\,secs]$. The error for these
estimated positions are computed with the actual positions. We also observed the number of
localization calls in $[0,100\,secs]$. The experiment is approximately repeated $10000$ times. The
data are grouped with respect to number of localization calls. In Figure~\ref{ErrEnrgFig}, we plot
the average error for different number of localization counts.
This figure shows that MAINT performs uniformly better than MADRD. For fixed error level, the
localization count and hence the energy consumption in MAINT is significantly lower. Also, for
comparable numbers of localization count, MAINT has much lower average error. It estimates the
position of a sensor with less error and even consuming less energy. MAINT locates a mobile sensor
with nearly exact position of the sensor consuming approximately half energy than that of MADRD.
MAINT requires higher memory to hold the history of location queries. We can hold limited number of
most recent query points. However, MAINT saves the valuable energy at the cost of cheap
memory.

\begin{figure}[h!] \vspace*{-5mm}
\centering
\input{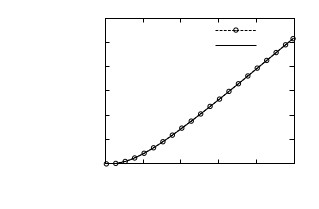}
\caption{Average error in estimation by the proposed technique and theoretical analysis.}
 \label{Sim:ErrTheo} \vspace*{-2.5mm}
\end{figure}

\begin{figure}[h!]  \vspace*{-5mm}
\centering
\input{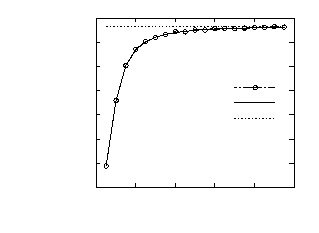}
\caption{Average error in estimation by the proposed
technique with $\frac{T}{\lambda}=50.0$.}
 \label{Sim:ErrVarT}  \vspace*{-2.5mm}
\end{figure}

In Figure~\ref{Sim:ErrTheo}, we compare expected error in position estimation using MAINT with an
expression.
This simulation was carried out under the same RWP model in C++ environment. In the course of this
simulation, MAINT calls localization procedure with fixed time period $T$. This process is repeated
at least $100$ times for a particular value of $T$. The average error are plotted with respect to
several values of $T$  in Figure~\ref{Sim:ErrTheo}. This shows that error may be computed from the
deduced expression. Simulation studies with $\frac{T}{\lambda}=50.0$ and $\sigma = 10.0$ are shown
in the Figure~\ref{Sim:ErrVarT}.
It shows the asymptotic nature of the average error. This plotting shows that the average error
becomes stable, if $T$ varies with $\frac{T}{\lambda}=constant$. However, from
Figure~\ref{Sim:ErrTheo} and \ref{Sim:ErrVarT}, we observe that our theoretical analysis well
supported by the simulation studies.

\section{Conclusion} \label{conclude} 

The technique, proposed in this paper, estimates the location of a mobile sensor. Tilak et al.
\cite{TKAK05} proposed MADRD, which uses extrapolation and the position of the sensor is estimated
by the velocity in between last two localization points. In our proposed method MAINT, we use
interpolation. The velocity is calculated from the last and next localization points. In the
simulation studies, we see that MAINT estimates the position of the sensor with much lower error
than that of MADRD. If the parameters of the model are known, at any moment, the error in the
position estimation may be computed from the deduced expression, instead using actual position.

The time interval can control the energy dissipation. A constant error limit can be maintained if
the time period of localization increases proportionally to the rate of change of direction of its
motion. Increasing time period, the energy may be saved with a stable error limit. From analysis, we
observe that when a sensor changes the direction in its motion, our proposed technique provides
location with very low error as oppose to the methods proposed by Tilak et al.

Work is in progress to analyze the performances of the proposed model under other  movement models
like the Gaussian movement model, Brownian motion model etc.

\bibliographystyle{plain}
\bibliography{../../../mybibfile}

\end{document}